\documentclass[11pt,aps,showpacs,nofootinbib,prd,aps,epsf,floats,
               amsmath,amssymb,amsfonts]{revtex4}
\usepackage{amsmath, amssymb} 

\newcommand{\mathsym}[1]{{}} 

\usepackage{graphicx}% Include figure files
\usepackage{amsmath} \usepackage{amssymb}
\usepackage{bm}% bold math
\newcommand{\be}{\begin{equation}} \newcommand{\ee}{\end{equation}} \newcommand{\bea}{\begin{eqnarray}}
\newcommand{\eea}{\end{eqnarray}}  
\newcommand{\rem}[1]{}
\newsavebox{\PSLASH}
 \sbox{\PSLASH}{$p$\hspace{-1.8mm}/}
 
\renewcommand{\theequation}{\thesection.\arabic{equation}} \newcounter{saveeqn}
\newcommand{\add}{\addtocounter{equation}{1}}
\newcommand{\alpheqn}{\setcounter{saveeqn}{\value{equation}}%
\setcounter{equation}{0}%
\renewcommand{\theequation}{\mbox{\thesection.\arabic{saveeqn}{\alph{equation}}}}}
\newcommand{\reseteqn}{\setcounter{equation}{\value{saveeqn}}%
\renewcommand{\theequation}{\thesection.\arabic{equation}}}

 \newsavebox{\notrightarrow}
 \sbox{\notrightarrow}{$\to$\hspace{-4mm}/}
 
 \newsavebox{\PARTIALSLASH}
 \sbox{\PARTIALSLASH}{$\partial$\hspace{-1.6mm}/}
 
 \newsavebox{\ASLASH}
 \sbox{\ASLASH}{$A$\hspace{-2.1mm}/}
 
 \newsavebox{\KSLASH}
 \sbox{\KSLASH}{$k$\hspace{-1.8mm}/}
 
 \newsavebox{\LSLASH}
 \sbox{\LSLASH}{$\ell$\hspace{-1.8mm}/}
 
 \newsavebox{\QSLASH}
 \sbox{\QSLASH}{$q$\hspace{-1.8mm}/}
 
 \newsavebox{\DSLASH}
 \sbox{\DSLASH}{$D$\hspace{-2.2mm}/}
 
 \newsavebox{\DbfSLASH}
 \sbox{\DbfSLASH}{${\mathbf D}$\hspace{-2.8mm}/}
 
 \newsavebox{\DELVECRIGHT}
 \sbox{\DELVECRIGHT}{$\stackrel{\rightarrow}{\partial}$}
 
 \newcommand{\blue}{\IfColor{\textCadetBlue}{}}
\newcommand{\black}{\IfColor{\textBlack}{}} \newcommand{\red}{\IfColor{\textRed}{}}
\newcommand{\green}{\IfColor{\textOliveGreen}{}} \newcommand{\lila}{\IfColor{\textRedViolet}{}}

\begin{document}
\begin{center}
\centerline{{\bf\Large{Noncommutative Sugawara Construction}}} \vspace{6mm}
\bf{M. Ghasemkhani\footnote{e-mail: $m_{_{-}}$ghasemkhani@sbu.ac.ir}$^{;\ a}$}
\\
\normalsize
\bigskip\medskip
{$^a$ \it Department of Physics, Shahid Beheshti University, \\G.C., Evin, Tehran 19839, Iran}\\
 \end{center}
%%%%%%%%%%%%%%%%%%%%%%%%%%%%%%%%%%%%%
\begin{abstract}
\noindent
The noncommutative extension of the Sugawara construction for free massless fermionic
fields in two dimensions is studied. We prove that the equivalence of the
noncommutative Sugawara energy-momentum tensor and symmetric energy-momentum tensor
persists in the noncommutative extension. Some relevant physical results of this equivalence are also discussed.
\end{abstract}  \maketitle
%%%%%%%%%%%%%%%%%%%%%%%%
\section{Introduction}\label{introduction}
\noindent One of the outstanding features of
two dimensional field theories is bosonization where a free massless fermionic field can be written as a bosonic filed.
 This property
is rooted in the work of Jordan and Wigner \cite{jordan} where it was shown that the fermionic creation and
annihilation operators may be represented as the bosonic counterparts. On the other hand, the idea of
describing strong interaction process in terms of currents was proposed in \cite{gellmann,dashen,sugy}. In
this approach, the dynamical variables are taken to be the currents and the canonical formalism is abandoned.
In other words, each particle does not correspond to a field which satisfies the canonical commutation relation but Hilbert space is built upon current operators.
Accordingly, it was shown that the energy momentum-tensor of these theories can be expressed as a quadratic
function of the currents known as Sugawara construction \cite{sugy}. Later on, it was proved that the
symmetric energy-momentum tensor of the two-dimensional free massless fermionic theory is exactly equivalent
to the Sugawara energy-momentum tensor which is bilinear in fermionic currents \cite{gross}.\\
 \noindent
 Indeed, this equivalence confirmed the result of \cite{jordan}, equivalence of free massless fermions and
 bosons, in an elegant way. Then generalization of this famous equivalence to the curved space-time was
 performed in \cite{davies} where boson-fermion correspondence was shown for a general metric in two
 dimensions.\\
\noindent
Our purpose is to study whether this equivalence is satisfied for noncommutative space, where the
nature of the space-time changes at very short distances \cite{connes, sheikh,douglas}, which is not a trivial extension.
The authors in \cite{banerjee} considered the noncommutative generalization of the Sugawara energy-momentum tensor and then used the Seiberg-Witten map. While in the present work, the correspondence between the noncommutative Sugawara
construction and the symmetric energy-momentum tensor for two dimensional free fermionic theory is
addressed, without employing the Seiberg-Witten map. Applying the techniques described in \cite{gross}, we demonstrate that the noncommutative
Sugawara energy-momentum tensor exactly leads to the symmetric energy-momentum tensor.
An interesting physical consequence of this equivalence is noncommutative bosonization
that exhibits the relationship between the fermionic and bosonic fields in noncommutative space, as will be
discussed in the last section.
%%%%%%%%%%%
\section{Equivalence of the symmetric energy-momentum tensor and Sugawara energy-momentum tensor in
noncommutative space}
%%%%%%%%%%%%%%%%%%%%%%%%%%%%%%%%%%%%%%%%%%%%%%%%%%%%%%%%%%%
\noindent
The first part of this section includes derivation of the symmetric energy-momentum tensor using
variation of the action with respect to a generic metric. In the second part, we extend the Sugawara
construction to noncommutative space and demonstrate that it will be equivalent to the symmetric
energy-momentum tensor.
%%%%%%%%%%%%%%%%%%%%%%%%%%%%%%%%%%%%%%%%%%%%%%%%%%%%%%%%%%%%%%%%%%
\subsection{Symmetric energy-momentum tensor}
%%%%%%%%%%%%%%%%%%%%%%%%%%%%%%%%%%%%%%%%%%%%%%%%%%%%%%%%%%%%%%%%%%%
\noindent
 Let us start from the noncommutative version of the free massless fermionic Lagrangian density, which is obtained
 by replacing the ordinary product with the star-product
\begin{eqnarray}\label{lagrangian} {\cal
L}=\frac{i}{2}\left(\bar\psi\gamma^{\alpha}\star\partial_{\alpha}\psi
-\partial_{\alpha}\bar\psi\gamma^{\alpha}\star\psi\right),
\end{eqnarray}
where the star-product is defined as follows
\begin{eqnarray}\label{star-def} f(x)\star
g(x)\equiv \exp\left(\frac{i\theta_{\alpha\beta}}{2}\frac{\partial}{\partial
a_{\alpha}}\frac{\partial}{\partial b_{\beta}}\right) f(x+a)g(x+b)\Bigg|_{a,b=0},
\end{eqnarray}
 here
$\theta_{\mu\nu}$ is an antisymmetric constant matrix. As is usual in two-dimensional field theory, we choose to work in Euclidean
signature. In the noncommutative version, this has the added virtue that the Euclidean theory does not have issues with unitarity.
The noncommutative symmetric
energy-momentum tensor $T_{\mu\nu}^{^{\star}}$ is achieved by variation of the action $S$ with respect to a
generic metric $g_{\mu\nu}$ and setting $g^{\mu\nu}=\delta^{\mu\nu}$ in the end \cite{ashok}:
\begin{eqnarray}\label{def ofem tensor}
T_{\mu\nu}^{^{\star}}=\frac{2}{\sqrt{g}}\frac{\delta S}{\delta
g^{\mu\nu}(x)}\bigg|_{g^{\mu\nu}=\delta^{\mu\nu}},
 \end{eqnarray}
 where $g$ indicates the determinant of the metric with signature $(+,+)$.
The variation of the action corresponding to the Lagrangian density (\ref{lagrangian}) can be written as
 \begin{eqnarray}
\delta S&=&\frac{i}{8}\int
d^{2}y\sqrt{g}\star\left(\bar\psi\star\gamma_{\alpha}\delta g^{\alpha\beta}
\star\partial_{\beta}\psi+
\bar\psi\star\gamma_{\alpha}\delta g^{\beta\alpha}\star\partial_{\beta}\psi-
\partial_{\alpha}\bar\psi\star\gamma_{\beta}\delta g^{\alpha\beta}\star\psi
-\partial_{\alpha}\bar\psi\star\gamma_{\beta}\delta g^{\beta\alpha}\star\psi\right)\nonumber\\
&+&\frac{i}{2}\int
d^{2}y~\left(\delta\sqrt{g}\right)\star\left(\bar\psi\gamma^{\alpha}\star\partial_{\alpha}\psi
-\partial_{\alpha}\bar\psi\gamma^{\alpha}\star\psi\right).
 \end{eqnarray}
 Using the
relation (\ref{def ofem tensor}) and the cyclic property of the star product under the integral, we have
\begin{eqnarray}
T_{\mu\nu}^{^{\star}}&=&-\frac{i}{4} \left(\partial_{\nu}\psi_{\beta}\star
\bar\psi_{\alpha}(\gamma_{\mu})^{\alpha\beta}+\partial_{\mu}\psi_{\beta}\star
\bar\psi_{\alpha}(\gamma_{\nu})^{\alpha\beta}+\partial_{\mu}\bar\psi_{\alpha}\star\psi_{\beta}(\gamma_{\nu})^{\alpha\beta}
+\partial_{\nu}\bar\psi_{\alpha}\star\psi_{\beta}(\gamma_{\mu})^{\alpha\beta}\right)
\nonumber\\&-&\frac{i}{2}\delta^{\mu\nu}\left(
\bar\psi_{\alpha}\star\partial_{\lambda}\psi_{\beta}(\gamma^{\lambda})^{\alpha\beta}
-\partial_{\lambda}\bar\psi_{\alpha}\star\psi_{\beta}(\gamma^{\lambda})^{\alpha\beta}\right).
 \end{eqnarray}
Applying the equation of motion for free massless fermions, we find the energy-momentum tensor as
\begin{eqnarray}\label{final-canonical}
T_{\mu\nu}^{^{\star}}&=&-\frac{i}{4} \left(\partial_{\nu}\psi_{\beta}\star
\bar\psi_{\alpha}(\gamma_{\mu})^{\alpha\beta}+\partial_{\mu}\psi_{\beta}\star
\bar\psi_{\alpha}(\gamma_{\nu})^{\alpha\beta}+\partial_{\mu}\bar\psi_{\alpha}\star\psi_{\beta}
(\gamma_{\nu})^{\alpha\beta}
+\partial_{\nu}\bar\psi_{\alpha}\star\psi_{\beta}(\gamma_{\mu})^{\alpha\beta}\right),
 \end{eqnarray}
 which is
completely symmetric under $\mu\leftrightarrow\nu$.
%%%%%%%%%%%%%%%%%%%%%%%%%%%%%%%%%%%%%%%%%%%%%%%%%%%%%%%%%%%%%%%%%%%%%%%%%%%%%%%%%%%%%
\subsection{Sugawara energy-momentum tensor}
\noindent
The equivalence of the Sugawara construction and the symmetric energy-momentum tensor in commutative space has been shown in \cite{gross} and is reviewed in
appendix A. In the present section, we construct the noncommutative version of the Sugawara energy-momentum
tensor to demonstrate that it is precisely equivalent to (\ref{final-canonical}).\\ \noindent The Lagrangian
(\ref{lagrangian}) is invariant under global U(1) transformation which yields two different Noether currents
\cite{Hayakawa}
\begin{eqnarray}\label{currents}
{\cal
J}_{\mu}(x)=:\bar\psi_{\alpha}(x)\star\psi_{\beta}(x):(\gamma_{\mu})^{\alpha\beta},~~~~
J_{\mu}(x)=:\psi_{\beta}(x)\star\bar\psi_{\alpha}(x):(\gamma_{\mu})^{\alpha\beta},
\end{eqnarray}
where $:$ $:$ denotes normal
ordering. Now, we extend the commutative Sugawara construction to the noncommutative one as a bilinear
function of $J_{\mu}(x)$ with inserting star product instead of ordinary product
\begin{eqnarray}\label{nc-sugawara}
T_{\mu\nu}^{s^{\star}}=\frac{1}{2c}\bigg(~J_{\mu}(x)\star J_{\nu}(x)+
J_{\nu}(x)\star J_{\mu}(x)-\delta_{\mu\nu} J^{\lambda}(x)\star J_{\lambda}(x)\bigg),
\end{eqnarray}
 where $c$ is the Schwinger constant which appears in the equal-time commutator of currents. Since the mass
 dimension of energy-momentum tensor and of the currents in two dimensions is equal to two and one respectively,
 the coefficient $c$ should be dimensionless while in four dimensions, it is a dimensionful quantity
 with dimension of a mass
square. The
 detailed analysis of the current algebra in two dimensions shows that the value of $c$ in noncommutative
 case
 is the same as the commutative one, $c=\frac{1}{\pi}$ \cite{Ghezelbash}.
 To prove that (\ref{nc-sugawara}) is exactly equivalent to (\ref{final-canonical}), we need to regularize the operator products in (\ref{nc-sugawara}).
 To this end, we use
 the point-splitting technique \cite{schwinger-algebra} and replace $J_{\mu}(x)\star
 J_{\nu}(x)$
 with
\begin{eqnarray} &&\lim_{\epsilon\rightarrow 0}~\bigg(J_{\mu}(x+\epsilon)\star J_{\nu}(x)-\langle
J_{\mu}(x+\epsilon)\star J_{\nu}(x)\rangle\bigg),
\end{eqnarray}
 which leads to
 \begin{eqnarray}\label{nc-sugawara-1}
T_{\mu\nu}^{s^{\star}}&=&\frac{\pi}{2}\lim_{\epsilon\rightarrow 0}\bigg(J_{\mu}(x+\epsilon)\star
J_{\nu}(x)+ J_{\nu}(x+\epsilon)\star J_{\mu}(x)- \delta_{\mu\nu} J^{\lambda}(x+\epsilon)\star
J_{\lambda}(x)\nonumber\\ &-&\langle J_{\mu}(x+\epsilon)\star J_{\nu}(x) \rangle-\langle J_{\nu}(x+\epsilon)\star J_{\mu}(x)
\rangle+\delta_{\mu\nu}\langle J^{\lambda}(x+\epsilon)\star J_{\lambda}(x) \rangle \bigg).
\end{eqnarray}
To perform
some algebraic manipulations on (\ref{nc-sugawara-1}), we employ the star product definition (\ref{star-def})
\begin{eqnarray}\label{operator-definition}
 J_{\mu}(x+\epsilon)=
{\cal F}_{ab}:\psi_{\beta}(x+\epsilon+a)\bar\psi_{\alpha}(x+\epsilon+b):\Bigg|_{a,b=0}(\gamma_{\mu})^{\alpha\beta},
\end{eqnarray}
 where ${\cal F}_{ab}$ is an abbreviated notation for the exponential operator appearing in (\ref{star-def}).
Accordingly, the first term of the equation (\ref{nc-sugawara-1}) can be written as
\begin{eqnarray}\label{abbreviated-form}
 J_{\mu}(x+\epsilon)\star J_{\nu}(x)&=&
{\cal F}_{fg}{\cal F}_{ab}{\cal F}_{cd}:\psi_{\beta}(x+\epsilon+f+a)\bar\psi_{\alpha}(x+\epsilon+f+b)::\psi_{\sigma}
(x+g+c)\bar\psi_{\rho}(x+g+d):\nonumber\\ &\times
&(\gamma_{\mu})^{\alpha\beta}(\gamma_{\nu})^{\rho\sigma}\Bigg|_{f,g,a,b,c,d=0}.
\end{eqnarray}
 Using Wick's theorem, (\ref{abbreviated-form}) changes into
\begin{eqnarray} \label{simple-form}
 J_{\mu}(x+\epsilon)\star
 J_{\nu}(x)&=&{\cal F}_{fg}{\cal F}_{ab}{\cal F}_{cd}\bigg(:\psi_{\beta}(x+\epsilon+f+a)\bar\psi_{\alpha}(x+\epsilon+f+b)\psi_{\sigma}(x+g+c)\bar\psi_{\rho}(x+g+d):\nonumber\\&-&
:\psi_{\sigma}(x+g+c)\langle\psi_{\beta}(x+\epsilon+f+a)\bar\psi_{\rho}(x+g+d)\rangle\bar\psi_{\alpha}(x+\epsilon+f+b):\nonumber\\&-&
:\psi_{\beta}(x+\epsilon+f+a)\langle\psi_{\sigma}(x+g+c)\bar\psi_{\alpha}(x+\epsilon+f+b) \rangle\bar\psi_{\rho}(x+g+d):
\nonumber\\&-& \langle\psi_{\beta}(x+\epsilon+f+a)\bar\psi_{\rho}(x+g+d)
\rangle\langle\psi_{\sigma}(x+g+c)\bar\psi_{\alpha}(x+\epsilon+f+b)\rangle\bigg)\nonumber\\ &\times
&(\gamma_{\mu})^{\alpha\beta}(\gamma_{\nu})^{\rho\sigma}\Bigg|_{f,g,a,b,c,d=0}.
 \end{eqnarray}
  Rewriting the
other terms of (\ref{nc-sugawara-1}) similar to (\ref{simple-form}) and substituting them again into
(\ref{nc-sugawara-1}), we obtain
\begin{eqnarray}\label{final form}
T_{\mu\nu}^{s^{\star}}=\frac{\pi}{2}\lim_{\epsilon\rightarrow 0}\bigg[{\cal Q}_{\mu\nu}(x,\epsilon)-{\cal
R}_{\mu\nu}(x,\epsilon)-{\cal R}_{\nu\mu}(x,\epsilon)-{\cal S}_{\nu\mu}(x,-\epsilon)-{\cal
S}_{\mu\nu}(x,-\epsilon)-\delta_{\mu\nu}[{\cal R}_{\lambda}^{\lambda}(x,\epsilon)+{\cal
S}_{\lambda}^{\lambda}(x,-\epsilon)]\bigg],\nonumber\\
\end{eqnarray}
 with
\begin{eqnarray}\label{components-of-T} {\cal
Q}_{\mu\nu}(x,\epsilon)&=&{\cal F}_{fg}{\cal F}_{ab}{\cal F}_{cd}:\bigg(\psi_{\beta}(x+\epsilon+f+a)\bar\psi_{\alpha}(x+\epsilon+f+b)\psi_{\sigma}
(x+g+c)\bar\psi_{\rho}(x+g+d)(\gamma_{\mu})^{\alpha\beta}(\gamma_{\nu})^{\rho\sigma}\nonumber\\&+&
\psi_{\beta}(x+\epsilon+f+a)\bar\psi_{\alpha}(x+\epsilon+f+b)\psi_{\sigma}
(x+g+c)\bar\psi_{\rho}(x+g+d)(\gamma_{\nu})^{\alpha\beta}(\gamma_{\mu})^{\rho\sigma}\nonumber\\
&-&\psi_{\beta}(x+\epsilon+f+a)\bar\psi_{\alpha}(x+\epsilon+f+b)\psi_{\sigma}
(x+g+c)\bar\psi_{\rho}(x+g+d)\delta_{\mu\nu}\nonumber\\&\times&(\gamma_{\lambda})^{\alpha\beta}(\gamma^{\lambda})^{\rho\sigma}\bigg):\Bigg|_{f,g,a,b,c,d=0},
\nonumber\\
 {\cal
R}_{\mu\nu}(x,\epsilon)&=&{\cal F}_{fg}{\cal F}_{ab}{\cal F}_{cd}:\psi_{\sigma}(x+g+c)\langle\psi_{\beta}(x+\epsilon+f+a)\bar\psi_{\rho}(x+g+d)
\rangle\bar\psi_{\alpha}(x+\epsilon+f+b):\nonumber\\
&\times&(\gamma_{\mu})^{\alpha\beta}(\gamma_{\nu})^{\rho\sigma}\Bigg|_{f,g,a,b,c,d=0},\nonumber\\
 {\cal
S}_{\nu\mu}(x,-\epsilon)&=&{\cal F}_{fg}{\cal F}_{ab}{\cal F}_{cd}:\psi_{\beta}(x+\epsilon+f+a)\langle\psi_{\sigma}(x+g+c)\bar\psi_{\alpha}(x+\epsilon+f+b)
\rangle\bar\psi_{\rho}(x+g+d):\nonumber\\ &\times
&(\gamma_{\mu})^{\alpha\beta}(\gamma_{\nu})^{\rho\sigma}\Bigg|_{f,g,a,b,c,d=0}.
\end{eqnarray}
 The field ordering appearing in the vacuum expectation value of the relation (\ref{components-of-T}) is
 not time ordering and is defined as\footnote{ We follow the convention used in
 \cite{gross}.}
\begin{eqnarray}
 S^{(+)}(x-y)=\psi(x)\bar\psi(y)-:\psi(x)\bar\psi(y):=\langle\psi(x)\bar\psi(y)\rangle.
\end{eqnarray}
 In
view of the above definition, the quantities ${\cal R}_{\mu\nu}(x,\epsilon)$ and ${\cal
S}_{\nu\mu}(x,-\epsilon)$ can be represented as
\begin{eqnarray}\label{R-S}
 {\cal
R}_{\mu\nu}(x,\epsilon)&=&{\cal F}_{fg}{\cal F}_{ab}{\cal F}_{cd}:\psi_{\sigma}(x+g+c)
\bar\psi_{\alpha}(x+\epsilon+f+b):\nonumber\\
&\times&S^{(+)}_{\beta\rho}(\epsilon+f+a-g-d)(\gamma_{\mu})^{\alpha\beta}(\gamma_{\nu})^{\rho\sigma}\Bigg|_{f,g,a,b,c,d=0},\nonumber\\
{\cal
S}_{\nu\mu}(x,-\epsilon)&=&{\cal F}_{fg}{\cal F}_{ab}{\cal F}_{cd}:\psi_{\beta}(x+\epsilon+f+a)\bar\psi_{\rho}(x+g+d):\nonumber\\
&\times
&S^{(+)}_{\sigma\alpha}(g+c-\epsilon-f-b)(\gamma_{\mu})^{\alpha\beta}(\gamma_{\nu})^{\rho\sigma}\Bigg|_{f,g,a,b,c,d=0}.
\end{eqnarray}
Converting equation (\ref{R-S}) to the star product form, we obtain
\begin{eqnarray}
&&{\cal
R}_{\mu\nu}(x,\epsilon)=-\left(\gamma_{\mu}
S^{(+)}(\epsilon)\gamma_{\nu}\right)^{\alpha\beta}:\bar\psi_{\alpha}(x+\epsilon)
\star\psi_{\beta}(x):,\nonumber\\ &&{\cal S}_{\nu\mu}(x,-\epsilon)=\left(\gamma_{\nu}
S^{(+)}(-\epsilon)\gamma_{\mu}\right)^{\alpha\beta}:\psi_{\beta}(x+\epsilon) \star\bar\psi_{\alpha}(x):,
\end{eqnarray}
 where
 \begin{eqnarray} S^{(+)}(\epsilon)=
-\frac{i}{2\pi}\frac{\epsilon_{\xi}\gamma^{\xi}}{\epsilon^{2}}.
\end{eqnarray}
 Note that the minus sign in ${\cal
R}_{\mu\nu}(x,\epsilon)$ comes from the odd permutation of the fermionic fields.\\ Expanding $\psi$ and
$\bar\psi$ up to the first order in $\epsilon$ yields
\begin{eqnarray} &&{\cal
R}_{\mu\nu}(x,\epsilon)=\frac{i\epsilon^{\xi}}{2\pi\epsilon^{2}}\left(\gamma_{\mu}\gamma_{\xi}
\gamma_{\nu}\right)^{\alpha\beta}
:[\bar\psi_{\alpha}(x)+\epsilon^{\eta}\partial_{\eta}\bar\psi_{\alpha}(x)+{\cal O}(\epsilon^{2})]
\star\psi_{\beta}(x):, \nonumber\\ &&{\cal S}_{\nu\mu}(x,-\epsilon)=\frac{i\epsilon^{\xi}}{2\pi\epsilon^{2}}
\left(\gamma_{\nu}\gamma_{\xi}\gamma_{\mu}\right)^{\alpha\beta}
:[\psi_{\beta}(x)+\epsilon^{\eta}\partial_{\eta}\psi_{\beta}(x)+{\cal O}(\epsilon^{2})]
\star\bar\psi_{\alpha}(x):,
 \end{eqnarray}
 and using the following symmetric limits
\begin{eqnarray}\label{sym-lim}
\lim_{\epsilon\rightarrow 0}~(\frac{\epsilon^{\alpha}}{\epsilon^{2}})=0,~~~~~~~
\lim_{\epsilon\rightarrow 0}~(\frac{\epsilon^{\alpha}\epsilon^{\beta}}{\epsilon^{2}})=\frac{1}{2}\delta_{\alpha\beta},
\end{eqnarray}
we conclude that
\begin{eqnarray}\label{final-value-R-S}
&&\lim_{\epsilon\rightarrow 0}{\cal R}_{\mu\nu}(x,\epsilon)=\frac{i}{4\pi}
\left(\gamma_{\mu}\gamma_{\xi}\gamma_{\nu}\right)^{\alpha\beta} :\partial^{\xi}\bar\psi_{\alpha}(x)
\star\psi_{\beta}(x):, \nonumber\\ &&\lim_{\epsilon\rightarrow 0}{\cal
S}_{\nu\mu}(x,-\epsilon)=\frac{i}{4\pi} \left(\gamma_{\nu}\gamma_{\xi}\gamma_{\mu}\right)^{\alpha\beta}
:\partial^{\xi}\psi_{\beta}(x) \star\bar\psi_{\alpha}(x):.
\end{eqnarray}
 Inserting the result
(\ref{final-value-R-S}) in (\ref{final form}), we arrive at
 \begin{eqnarray}\label{final-form-1}
T_{\mu\nu}^{s^{\star}}&=&\frac{\pi}{2}{\cal Q}_{\mu\nu}(x)-\frac{i}{8}:\bigg[
\left(\gamma_{\mu}\gamma_{\xi}\gamma_{\nu}+\gamma_{\nu}\gamma_{\xi}\gamma_{\mu}\right)^{\alpha\beta}
\left(\partial^{\xi}\bar\psi_{\alpha}(x) \star\psi_{\beta}(x)+ \partial^{\xi}\psi_{\beta}(x)
\star\bar\psi_{\alpha}(x)\right)\nonumber\\&+&\delta_{\mu\nu}(\gamma^{\lambda}
\gamma_{\xi}\gamma_{\lambda})^{\alpha\beta}\left(\partial^{\xi}\bar\psi_{\alpha}(x)
\star\psi_{\beta}(x)+\partial^{\xi}\psi_{\beta}\star\bar\psi_{\alpha}\right)\bigg]:.
\end{eqnarray}
The product
of gamma matrices in (\ref{final-form-1}) can be simplified using the Clifford algebra
\begin{eqnarray}\label{identity}
\gamma_{\mu}\gamma_{\xi}\gamma_{\nu}+\gamma_{\nu}\gamma_{\xi}\gamma_{\mu}=
2\left(\delta_{\mu\xi}\gamma_{\nu}+\delta_{\nu\xi}\gamma_{\mu}-\delta_{\mu\nu} \gamma_{\xi}\right)
,~~~~~~~~~ \gamma_{\lambda}\gamma_{\xi}\gamma^{\lambda}=(2-d)\gamma_{\xi}.
\end{eqnarray}
Substituting
(\ref{identity}) into (\ref{final-form-1}) and applying the equation of motion
$\gamma^{\xi}\partial_{\xi}\psi=0$ and $\partial_{\xi}\bar\psi\gamma^{\xi}=0$, we obtain the Sugawara energy-momentum tensor
\begin{eqnarray}\label{final-form-2}
T_{\mu\nu}^{s^{\star}}&=&\frac{\pi}{2}{\cal
Q}_{\mu\nu}(x)-\frac{i}{4}:\bigg[ \bigg(\partial_{\nu}\bar\psi_{\alpha}(x)
\star\psi_{\beta}(x)+\partial_{\nu}\psi_{\beta}(x)\star\bar\psi_{\alpha}(x)\bigg)(\gamma_{\mu})^{\alpha\beta}
\nonumber\\&+& \bigg(\partial_{\mu}\bar\psi_{\alpha}(x)\star\psi_{\beta}(x)+
\partial_{\mu}\psi_{\beta}(x)\star\bar\psi_{\alpha}(x)\bigg)(\gamma_{\nu})^{\alpha\beta}\bigg]:.
 \end{eqnarray}
 We notice that the last term
of (\ref{final-form-1}) vanishes in two dimensions as a result of the identity
 $\gamma_{\lambda}\gamma_{\xi}\gamma^{\lambda}=(2-d)\gamma_{\xi}$.
  In order to show that $T_{\mu\nu}^{^{\star}}=T_{\mu\nu}^{s^{\star}}$, it is enough to
 demonstrate ${\cal{Q}}_{\mu\nu}=0$. For simplicity, we carry out computations
  in
the light-cone coordinate system, $x_{\pm} = x_{1} \pm ix_{2}$.
The representation of the Euclidean gamma matrices
  \begin{eqnarray}\label{Euclidean-gamma}
\gamma_{1}=\left(
  \begin{array}{cc}
    0 & -i \\
    i & 0 \\
  \end{array}
\right),~~~~\gamma_{2}=\left(
  \begin{array}{cc}
    0 & -1 \\
    -1 & 0 \\
  \end{array}\right),
  \end{eqnarray}
 in the light-cone coordinates is given by
  \begin{eqnarray}
\gamma_{+}=\gamma_{1}+i\gamma_{2}= \left(
  \begin{array}{cc}
    0 & -2i\\
   0 & 0 \\
  \end{array}\right)
,~~~~ \gamma_{-}=\gamma_{1}-i\gamma_{2}= \left(
  \begin{array}{cc}
    0 & 0\\
   2i & 0 \\
  \end{array}\right),
  \end{eqnarray}
with \begin{eqnarray} g^{\mu\nu}=\left(
  \begin{array}{cc}
    g^{++} & g^{+-}\\
   g^{-+} & g^{--} \\
  \end{array}\right)=
  \left(
  \begin{array}{cc}
    0 & \frac{1}{2}\\
 \frac{1}{2} & 0 \\
  \end{array}\right).
\end{eqnarray}
The equation of motion for two-dimensional
massless fermions, which is described
by $i\gamma^{\mu}\partial_{\mu}\psi=0$ with
 $ \psi=\left(
    \begin{array}{c}
      \psi_{1}\\
      \psi_{2}\\
    \end{array}
  \right)$, in the
  light-cone coordinate system reduces to
$\partial_{+}\psi_{1}=\partial_{-}\psi_{2}=0$. Thus
 \begin{eqnarray}
&&\psi_{1}=\psi_{1}(x_{-}),~~~~~\psi_{2}=\psi_{2}(x_{+}),\nonumber\\&&
\bar\psi_{1}=\bar\psi_{1}(x_{+}),~~~~~\bar\psi_{2}=\bar\psi_{2}(x_{-}).
\end{eqnarray}
Using $[x_{+},x_{-}]=2\theta$, the
expression $\psi_{\beta}(x)\star\bar\psi_{\alpha}(x)(\gamma_{\mu})^{\alpha\beta}$ with on-shell Dirac fermions is rewritten as
\begin{eqnarray}\label{simp-lc}
&&\psi_{\beta}\star\bar\psi_{\alpha}(\gamma_{+})_{\alpha\beta}=
-2i~\psi_{2}(x_{+})\star\bar\psi_{1}(x_{+})=-2i~\psi_{2}(x_{+})\bar\psi_{1}(x_{+}),\nonumber\\
&&\psi_{\beta}\star\bar\psi_{\alpha}(\gamma_{-})_{\alpha\beta}=
+2i~\psi_{1}(x_{-})\star\bar\psi_{2}(x_{-})=+2i~\psi_{1}(x_{-})\bar\psi_{2}(x_{-}).
 \end{eqnarray}
 With equation (\ref{simp-lc}), it would be possible to find all the components of ${\cal Q}_{\mu\nu}$. Since
there is no singularity in ${\cal Q}_{\mu\nu}(x,\epsilon)$, we have
\begin{eqnarray}
 \lim_{\epsilon\rightarrow 0}{\cal Q}_{\mu\nu}(x,\epsilon)={\cal Q}_{\mu\nu}(x),
\end{eqnarray}
 and hence
 \begin{eqnarray} {\cal
Q}_{\mu\nu}(x)&=&:\bigg(\psi_{\beta}(x)\star\bar\psi_{\alpha}(x)\bigg)\star\bigg(\psi_{\sigma}
(x)\star\bar\psi_{\rho}(x)\bigg):(\gamma_{\mu})^{\alpha\beta}(\gamma_{\nu})^{\rho\sigma}\nonumber\\&+&:
\bigg(\psi_{\beta}(x)\star\bar\psi_{\alpha}(x)\bigg)\star\bigg(\psi_{\sigma}
(x)\star\bar\psi_{\rho}(x)\bigg):(\gamma_{\nu})^{\alpha\beta}(\gamma_{\mu})^{\rho\sigma}\nonumber\\
&-&:\bigg(\psi_{\beta}(x)\star\bar\psi_{\alpha}(x)\bigg)\star\bigg(\psi_{\sigma}
(x)\star\bar\psi_{\rho}(x)\bigg):\delta_{\mu\nu}(\gamma_{\lambda})^{\alpha\beta}(\gamma^{\lambda})^{\rho\sigma}.
\end{eqnarray}
One may then readily show that
 \begin{eqnarray}
  {\cal
Q}_{++}&=&-8:\bigg(\psi_{2}(x_{+})\bar\psi_{1}(x_{+})\bigg)\star\bigg(\psi_{2}(x_{+})\bar\psi_{1}(x_{+})\bigg):\nonumber\\
 &=&-8:\psi_{2}(x_{+})\bar\psi_{1}(x_{+})\psi_{2}(x_{+})\bar\psi_{1}(x_{+}):.
\end{eqnarray}
 Performing some straightforward permutations, we get ${\cal Q}_{++}=0$. Similarly
 \begin{eqnarray} {\cal
Q}_{--}&=&-8:\bigg(\psi_{1}(x_{-})\bar\psi_{2}(x_{-})\bigg)\star\bigg(\psi_{1}(x_{-})\bar\psi_{2}(x_{-})\bigg):\nonumber\\
 &=&-8:\psi_{1}(x_{-})\bar\psi_{2}(x_{-})\psi_{1}(x_{-})\bar\psi_{2}(x_{-}):\nonumber\\
 &=&0.
\end{eqnarray}
 Also for off-diagonal components ${\cal Q}_{\pm\mp}$, we have
\begin{eqnarray}
{\cal Q}_{\pm\mp}(x)&=&:\bigg(\psi_{\beta}(x)\star\bar\psi_{\alpha}(x)\bigg)\star\bigg(\psi_{\sigma}
(x)\star\bar\psi_{\rho}(x)\bigg):(\gamma_{\pm})^{\alpha\beta}(\gamma_{\mp})^{\rho\sigma}\nonumber\\&+&:
\bigg(\psi_{\beta}(x)\star\bar\psi_{\alpha}(x)\bigg)\star\bigg(\psi_{\sigma}
(x)\star\bar\psi_{\rho}(x)\bigg):(\gamma_{\mp})^{\alpha\beta}(\gamma_{\pm})^{\rho\sigma}\nonumber\\
&-&:2\bigg(\psi_{\beta}(x)\star\bar\psi_{\alpha}(x)\bigg)\star\bigg(\psi_{\sigma}
(x)\star\bar\psi_{\rho}(x)\bigg):(\gamma_{\lambda})^{\alpha\beta}(\gamma^{\lambda})^{\rho\sigma}.
\end{eqnarray}
 Inserting $\gamma^{\pm}=\frac{1}{2}\gamma_{\mp}$ results in
 \begin{eqnarray}
{\cal Q}_{\pm\mp}(x)&=& :\bigg(\psi_{\beta}(x)\star\bar\psi_{\alpha}(x)\bigg)\star\bigg(\psi_{\sigma}
(x)\star\bar\psi_{\rho}(x)\bigg):(\gamma_{\pm})^{\alpha\beta}(\gamma_{\mp})^{\rho\sigma}\nonumber\\&+&:
\bigg(\psi_{\beta}(x)\star\bar\psi_{\alpha}(x)\bigg)\star\bigg(\psi_{\sigma}
(x)\star\bar\psi_{\rho}(x)\bigg):(\gamma_{\mp})^{\alpha\beta}(\gamma_{\pm})^{\rho\sigma}\nonumber\\
&-&:\bigg(\psi_{\beta}(x)\star\bar\psi_{\alpha}(x)\bigg)\star\bigg(\psi_{\sigma}
(x)\star\bar\psi_{\rho}(x)\bigg):(\gamma_{\pm})^{\alpha\beta}(\gamma_{\mp})^{\rho\sigma}\nonumber\\
&-&:\bigg(\psi_{\beta}(x)\star\bar\psi_{\alpha}(x)\bigg)\star\bigg(\psi_{\sigma}
(x)\star\bar\psi_{\rho}(x)\bigg):(\gamma_{\mp})^{\alpha\beta}(\gamma_{\pm})^{\rho\sigma}\nonumber\\&=&0.
\end{eqnarray}
 Consequently $Q_{\mu\nu}=0$. This means that the equivalence of the Sugawara energy-momentum tensor and
  energy-momentum tensor $T_{\mu\nu}^{s^{\star}}=T_{\mu\nu}^{^{\star}}$ in two-dimensional
 noncommutative space for free massless fermions is still satisfied. Also this equivalence occurs for the Sugawara form in terms of the current
 ${\cal J}_{\mu}(x)$, as defined in (\ref{currents}). We have
  \begin{eqnarray}\label{sugawara-with-second-current}
\widehat{T}_{\mu\nu}^{s^{\star}}=\frac{\pi}{2}\bigg(~{\cal J}_{\mu}(x)\star {\cal J}_{\nu}(x)+
{\cal J}_{\nu}(x)\star {\cal J}_{\mu}(x)-\delta_{\mu\nu} {\cal J}^{\lambda}(x)\star {\cal J}_{\lambda}(x)\bigg).
\end{eqnarray}
To show this, let us write first all the components of the currents $J_{\mu}(x)$ and ${\cal J}_{\mu}(x)$ using the representation of the gamma matrices (\ref{Euclidean-gamma}) as follows
\begin{eqnarray}\label{current-1}
&&J_{1}(x)=i:\bigg(
\psi_{1}(x_{-})\star\bar\psi_{2}(x_{-})-\psi_{2}(x_{+})\star\bar\psi_{1}(x_{+})\bigg):=i:\bigg(
\psi_{1}(x_{-})\bar\psi_{2}(x_{-})-\psi_{2}(x_{+})\bar\psi_{1}(x_{+})\bigg):,\nonumber\\
&&J_{2}(x)=-:\bigg(
\psi_{2}(x_{+})\star\bar\psi_{1}(x_{+})+\psi_{1}(x_{-})\star\bar\psi_{2}(x_{-})\bigg):=-:\bigg(
\psi_{2}(x_{+})\bar\psi_{1}(x_{+})+\psi_{1}(x_{-})\bar\psi_{2}(x_{-})\bigg):,\nonumber\\
\end{eqnarray}
as well as
\begin{eqnarray}\label{current-2}
&&{\cal
J}_{1}(x)=i:\bigg(
\bar\psi_{2}(x_{-})\star\psi_{1}(x_{-})-\bar\psi_{1}(x_{+})\star\psi_{2}(x_{+})\bigg):=i:\bigg(
\bar\psi_{2}(x_{-})\psi_{1}(x_{-})-\bar\psi_{1}(x_{+})\psi_{2}(x_{+})\bigg):,\nonumber\\
&&{\cal
J}_{2}(x)=-:\bigg(
\bar\psi_{1}(x_{+})\star\psi_{2}(x_{+})+\bar\psi_{2}(x_{-})\star\psi_{1}(x_{-})\bigg):=-:\bigg(
\bar\psi_{1}(x_{+})\psi_{2}(x_{+})+\bar\psi_{2}(x_{-})\psi_{1}(x_{-})\bigg):.\nonumber\\
\end{eqnarray}
We notice that the star product appearing in the noncommutative currents is removed.
Applying the permutation on the fermionic fields of the relation (\ref{current-1}), we obtain
\begin{eqnarray}\label{equality-of-two-currents}
{\cal J}_{\mu}(x)=-J_{\mu}(x).
\end{eqnarray}
This is an interesting result in two dimensions.  Unlike the four-dimensional case, where $J_{\mu}(x)$ and ${\cal J}_{\mu}(x)$ correspond to each other by the charge conjugation transformation \cite{Jabbari}, which is not conserved, in two dimensions the charge conjugation, as well as the Lorentz invariance, remain the symmetry of the theory as in their  commutative case.
Inserting (\ref{equality-of-two-currents}) in (\ref{sugawara-with-second-current}) then leads to
 \begin{eqnarray}
\widehat{T}_{\mu\nu}^{s^{\star}}=T_{\mu\nu}^{s^{\star}}=T_{\mu\nu}^{^{\star}}~.
\end{eqnarray}
One of the physical consequences of this equivalence is noncommutative bosonization, which is obtained by writing the transformation of the field $\psi$ under the spatial translation
\begin{eqnarray}
\partial_{x_{1}}\psi(x)=i\bigg[P_{1},\psi(x)\bigg],~~~~~~~P_{1}=\int dx'_{1}T_{21}^{s^{\star}},
\end{eqnarray}
where $T_{21}^{s^{\star}}$ is the conserved current arising from translational invariance\footnote{In two-dimensional Euclidean space $x_{2}=ix_{0}$.}. We have
\begin{eqnarray}\label{nch}
\partial_{x_{1}}\psi(x)=i\bigg[\int dx_{1}'T_{21}^{s^{\star}},\psi(x)\bigg]_{x_{2}=x'_{2}},
\end{eqnarray}
and substituting the value of $T_{21}^{s^{\star}}$ from (\ref{nc-sugawara}) in (\ref{nch}) yields
\begin{eqnarray}\label{nc-Heisenberg}
\partial_{x_{1}}\psi(x)=\frac{i\pi}{2}\bigg[\int dx_{1}'\bigg(J_{1}(x')\star J_{2}(x')+J_{2}(x')\star J_{1}(x')\bigg),\psi(x)\bigg]_{x_{2}=x'_{2}}.
\end{eqnarray}
To simplify (\ref{nc-Heisenberg}),
we insert $\int dx'_{2}\delta(x_{2}-x'_{2})=1$ to use the trace property of the star product, which is given by
\begin{eqnarray}\label{trace-property}
\int dx'_{1}dx'_{2}~J_{2}(x')\star J_{1}(x')=\int dx'_{1}dx'_{2}~J_{1}(x')\star J_{2}(x').
\end{eqnarray}
Inserting (\ref{trace-property}) in (\ref{nc-Heisenberg}) and applying
 the operator definition of the star product (\ref{operator-definition}), we have
\begin{eqnarray}\label{nc-Heisenberg-2}
\partial_{x_{1}}\psi(x)&=&i\pi\int dx'_{1}dx'_{2}~\delta(x_{2}-x'_{2}){\cal F}_{ab}\bigg[J_{1}(x'+a)J_{2}(x'+b),\psi(x)\bigg]\Bigg|_{a,b=0}\nonumber\\&=&
i\pi\int dx'_{1}dx'_{2}~\delta(x_{2}-x'_{2}){\cal F}_{ab}\bigg(J_{1}(x'+a)\bigg[J_{2}(x'+b),\psi(x)\bigg]+\bigg[J_{1}(x'+a),\psi(x)\bigg]J_{2}(x'+b)
\bigg)\Bigg|_{a,b=0}.\nonumber\\
\end{eqnarray}
The bracket terms appearing in the right-hand side of (\ref{nc-Heisenberg-2}) are derived by considering the quantization condition $\{\psi_{\alpha}(x),\psi^{\dagger}_{\beta}(x')\}=\delta_{\alpha\beta}\delta
(x_{1}-x'_{1})$ as follows
\begin{eqnarray}\label{commutator-1}
&&\bigg[J_{2}(x'+a),\psi(x)\bigg]_{x_{2}=x'_{2}}=\psi(x'+a)\delta(x_{1}-x'_{1}+a),\nonumber\\
&&\bigg[J_{1}(x'+a),\psi(x)\bigg]_{x_{2}=x'_{2}}=\gamma_{5}\psi(x'+a)\delta(x_{1}-x'_{1}+a),
\end{eqnarray}
with $\gamma_{5}=i\gamma_{1}\gamma_{2}$. Substituting (\ref{commutator-1}) into (\ref{nc-Heisenberg-2}) and then converting the result into the star product form, we have
\begin{eqnarray}
\partial_{x_{1}}\psi(x)=
i\pi\bigg(J_{1}(x)+J_{2}(x)\gamma_{5}\bigg)\star\psi(x).
\end{eqnarray}
The solution of this equation is represented by
\begin{eqnarray}
\psi(x)={\cal P}\bigg(e^{i\pi\int_{-\infty}^{x_{1}} dx'_{1}[J_{1}(x')+J_{2}(x')\gamma_{5}]}_{\star}\bigg)\psi_{0},
\end{eqnarray}
where ${\cal P}$ denotes the path-ordering operator and $\psi_{0}$ is a constant spinor in two dimensions.
\\Now, as a result of (\ref{current-1}), we can use the bosonized form of the commutative current, which is introduced in appendix A. Hence, we conclude
\begin{eqnarray}
\psi(x)={\cal P}\bigg(e^{-i\sqrt{\pi}~[\gamma_{5}\phi(x)-\int_{-\infty}^{x_{1}} dx'_{1}\dot{\phi}(x')]}_{\star}\bigg)\psi_{0},
\end{eqnarray}
here $\dot{\phi}=\partial_{x'_{2}}\phi$.
 %%%%%%%%%%%%%%%%%%%%%%%%%%%%%%%%%%%%%%%%%%%%%%%%
\section{Discussion}
%%%%%%%%%%%%%%%%%%%%%%%%%%%%%%%%%%%%%%%%%%%%%%%%
\noindent
In this paper, we established the noncommutative extension of the Sugawara construction in bilinear form of the
currents for free massless fermions in two dimensions. It was shown that this construction
is precisely equivalent to the symmetric energy-momentum tensor.\\
\noindent
To prove the correctness of this equivalence, we determined the energy-momentum tensor in two separate methods. The first was the direct calculation using the symmetric definition of the energy-momentum tensor for on-shell Dirac fermions and
 the second contained a detailed analysis of noncommutative Sugawara construction by applying the point-splitting regularization. Furthermore, for simplification in our calculation, we considered the light-cone system. In this coordinate, we realized that the currents $J_{\mu}$ and ${\cal J}_{\mu}$, apart from a minus sign, are actually the same in two dimensions which leads to the charge conjugation symmetry restoration.
 \par
 Eventually, we presented a physical consequence of this equivalence, named as noncommutative bosonization (e.g. see \cite{nunez,grisaru,blas}), that relates a fermionic field to a bosonic field through an exponential function and demonstrated that a free massless fermion theory with a global $U(1)$ symmetry in noncommutative space corresponds to a free massless boson theory. Also, the bosonized version of a theory with local $U(1)$ symmetry such as two-dimensional noncommutative QED (NC-QED$_{2}$) was addressed in \cite{masoumeh}, where it was proven that
 the bosonized action contains a noncommutative Wess-Zumino-Witten (WZW) part, a gauge kinetic part and an interaction part between the WZW and gauge field. \newpage
 \noindent
 The physical significance of the bosonization procedure is that it specifies
   a duality between the strong and weak couplings for particular interacting quantum field theories.
       The most famous example of this duality is the equivalence of the massive Thirring model and sine-Gordon model \cite{Coleman,Mandelstam},
        where the weak coupling $\beta$ of the bosonic theory, that is the sine-Gordon model, is related to the strong coupling $g$ of the fermionic theory, the massive Thirring model, through the bosonization rule described by $\frac{4\pi}{\beta^{2}}=1+\frac{g}{\pi}$. Moreover, the duality between the noncommutative version of these models was studied in \cite{nunez,grisaru,blas} where it was shown that the strong-weak duality is also preserved.
         However, it is notable that the strong-weak duality does not appear in the case of NC-QED$_{2}$ and its bosonized version,
   because of the appearance of the same couplings in two theories.
%%%%%%%%%%%%%%%%%%%%%%%%%%%%%%%%%%%%%%%%%%%%%%%%
\section{Acknowledgments}
\noindent
I am grateful to M.M. Sheikh-Jabbari for numerous fruitful discussions and careful reading of the manuscript.
I would also like to thank M. Khorrami for his valuable comments and I appreciate the insightful and constructive remarks of M. Chaichian and the referee, which led to improvement of the manuscript.
Moreover, I acknowledge the School of Physics of Institute for research in fundamental sciences (IPM) for the research facilities and environment.
 %%%%%%%%%%%%%%%%%%%%%%%%%%%%%%%%%%%%%%%%%%%%%%%%
\par\noindent \begin{appendix}
%%%%%%%%%%%%%%%%%%%%%%%%%%%%%%%%%%%%%%%%%%%%%%%%%%%%%%%%%%%%%%%%%%
\section{Commutative Sugawara Construction} \setcounter{equation}{0}\noindent
%%%%%%%%%%%%%%%%%%%%%%%%%%%%%%%%%%%%%%%%%%%%%%%%%%%%%%%%%%%%%%%%%%%
In this appendix, we present a detailed analysis on the proof of the relation $T_{\mu\nu}^{s}=T_{\mu\nu}$
in two dimensional commutative space for free massless fermions, as argued in \cite{gross}, and describe some
interesting consequences of this equivalence \cite{Freundlich-1,Freundlich-2,sommerfield}. The Lagrangian for
the massless fermions is given by
\begin{eqnarray}
 {\cal
L}=\frac{i}{2}\left(\bar\psi\gamma^{\mu}\partial_{\mu}\psi-\partial_{\mu}\bar\psi\gamma^{\mu}\psi\right),
\end{eqnarray}
 which is invariant under the global phase transformation $\psi\rightarrow e^{i\alpha}\psi$ and
$\bar\psi\rightarrow e^{-i\alpha}\bar\psi$ that gives the conserved current
\begin{eqnarray}
j_{\mu}(x)=:\bar\psi(x)\gamma_{\mu}\psi(x):.
\end{eqnarray}
For this theory, the symmetric
energy-momentum tensor is written as follows
\begin{eqnarray}\label{com-canonic}
T_{\mu\nu}=\frac{i}{4}:\left(\bar\psi\gamma_{\mu}\partial_{\nu}\psi+\bar\psi\gamma_{\nu}\partial_{\mu}\psi-\partial_{\mu}
\bar\psi\gamma_{\nu}\psi-\partial_{\nu}\bar\psi\gamma_{\mu}\psi\right):.
 \end{eqnarray}
 The energy-momentum tensor in Sugawara form is described by a bilinear function of the currents as
\begin{eqnarray}\label{commutative-sugawara}
T_{\mu\nu}^{s}=\frac{\pi}{2}\left(~j_{\mu}(x)j_{\nu}(x)+j_{\nu}(x)j_{\mu}(x)-g_{\mu\nu}
j^{\lambda}(x)j_{\lambda}(x)\right).
\end{eqnarray}
 To show $T_{\mu\nu}^{s}=T_{\mu\nu}$, we start with
(\ref{commutative-sugawara}) and replace $j_{\mu}(x)j_{\nu}(x)$ with
\begin{eqnarray}\label{remove-divergency}
&&\lim_{\epsilon\rightarrow 0}~\bigg(j_{\mu}(x+\epsilon)j_{\nu}(x)-\langle j_{\mu}(x+\epsilon)j_{\nu}(x)\rangle\bigg).
\end{eqnarray}
Applying Wick's theorem on (\ref{remove-divergency})
\begin{eqnarray}
j_{\mu}(x+\epsilon)j_{\nu}(x)&=&:\bar\psi(x+\epsilon)\gamma_{\mu}\psi(x+\epsilon)::
\bar\psi(x)\gamma_{\nu}\psi(x):\nonumber\\&
=&:\bar\psi(x+\epsilon)\gamma_{\mu}\psi(x+\epsilon)\bar\psi(x)\gamma_{\nu}\psi(x):\nonumber\\&
+&:\bar\psi(x+\epsilon)\gamma_{\mu}\langle\psi(x+\epsilon)\bar\psi(x)\rangle\gamma_{\nu}\psi(x):\nonumber\\&
+&:\bar\psi(x)\gamma_{\nu}\langle\psi(x)\bar\psi(x+\epsilon)\rangle\gamma_{\mu}\psi(x+\epsilon):\nonumber\\&
-&tr\left(\gamma_{\mu}\langle\psi(x)\bar\psi(x+\epsilon)\rangle\gamma_{\nu}\langle\psi(x+\epsilon) \bar\psi(x)\rangle\right),
\end{eqnarray}
and implementing a similar analysis for the other terms of (\ref{commutative-sugawara}), we
arrive at
\begin{eqnarray}\label{sugawara in com} T_{\mu\nu}^{s}&=&\frac{\pi}{2}\lim_{\epsilon\rightarrow
0}\bigg[{\cal M}_{\mu\nu}(x,\epsilon)+{\cal{N}}_{\mu\nu}(x,\epsilon)+
{\cal{N}}_{\nu\mu}(x,\epsilon)+{\cal{N}}_{\mu\nu}(x,-\epsilon)+{\cal{N}}_{\nu\mu}(x,-\epsilon)\nonumber\\&-&g_{\mu\nu}[
{\cal{N}}_{\lambda}^{\lambda}(x,\epsilon)+{\cal{N}}_{\lambda}^{\lambda}(x,-\epsilon)]\bigg],\nonumber\\
\end{eqnarray}
 where ${\cal M}_{\mu\nu}$ and ${\cal N}_{\mu\nu}$ are defined as
 \begin{eqnarray} {\cal
M}_{\mu\nu}(x,\epsilon)&=&:\bar\psi(x+\epsilon)\gamma_{\mu}\psi(x+\epsilon)\bar\psi(x)\gamma_{\nu}\psi(x):\nonumber\\
&+& :\bar\psi(x+\epsilon)\gamma_{\nu}\psi(x+\epsilon)\bar\psi(x)\gamma_{\mu}\psi(x):\nonumber\\
&-&:\bar\psi(x+\epsilon)\gamma_{\lambda}\psi(x+\epsilon)\bar\psi(x)\gamma^{\lambda}\psi(x):g_{\mu\nu},
\nonumber\\ {\cal{N}}_{\mu\nu}(x,\epsilon)&=&:\bar\psi(x+\epsilon)\gamma_{\mu}S^{(+)}(\epsilon)
\gamma_{\nu}\psi(x):,
 \end{eqnarray}
and we have
 \begin{eqnarray}
S^{(+)}(\epsilon)=\langle\psi(x+\epsilon)\bar\psi(x)\rangle=
-(\frac{i}{2\pi})\frac{\epsilon_{\alpha}\gamma^{\alpha}}{\epsilon^{2}}.
\end{eqnarray}
 First, we concentrate on determining the value of ${\cal M}_{\mu\nu}(x)$.
\begin{eqnarray}
{\cal M}_{00}={\cal M}_{11}=:(j_{0})^{2}+(j_{1})^{2}:,~~~~~{\cal
M}_{01}={\cal M}_{10}=:j_{0}j_{1}+j_{1}j_{0}:.
\end{eqnarray}
 Choosing
$\gamma_{0}=\sigma_{z}$ and $\gamma_{1}=i\sigma_{y}$, we find
\begin{eqnarray}
j_{0}=\bar\psi_{1}\psi_{1}-\bar\psi_{2}\psi_{2},~~~~~j_{1}=\bar\psi_{1}\psi_{2}-\bar\psi_{2}\psi_{1}.
\end{eqnarray}
Therefore, all the components of ${\cal M}_{\mu\nu}$ in terms of the fermionic fields are given by
\begin{eqnarray} &&{\cal M}_{00}={\cal
M}_{11}=:(\bar\psi_{1}\psi_{1}-\bar\psi_{2}\psi_{2})^{2}+(\bar\psi_{1}
\psi_{2}-\bar\psi_{2}\psi_{1})^{2}:,\nonumber\\ &&{\cal M}_{01}={\cal
M}_{10}=:(\bar\psi_{1}\psi_{1}-\bar\psi_{2}\psi_{2})(\bar\psi_{1}\psi_{2}-\bar\psi_{2}\psi_{1})+(\bar\psi_{1}\psi_{2}-\bar\psi_{2}\psi_{1})
(\bar\psi_{1}\psi_{1}-\bar\psi_{2}\psi_{2}):.
 \end{eqnarray}
 Performing some permutations on fermionic fields yields ${\cal M}_{\mu\nu}=0$.
  In the next step, our purpose is to obtain the value of
    ${\cal N}_{\mu\nu}$. To this end, let us start
with expansion of the fermionic fields up to the first order in $\epsilon$
 \begin{eqnarray}
  &&{\cal
N}_{\mu\nu}(x,\epsilon)=-\frac{i\epsilon^{\xi}}{2\pi\epsilon^{2}}
:[\bar\psi(x)+\epsilon^{\alpha}\partial_{\alpha}\bar\psi(x)+{\cal O}(\epsilon^{2})]\gamma_{\mu}\gamma_{\xi}
\gamma_{\nu} \psi(x):
\end{eqnarray}
Taking the symmetric limits (\ref{sym-lim}),
\begin{eqnarray}\label{final-J}
&&\lim_{\epsilon\rightarrow 0}{\cal N}_{\mu\nu}(x,\epsilon)=\frac{i}{4\pi}
:\partial^{\xi}\bar\psi(x)\gamma_{\mu}\gamma_{\xi} \gamma_{\nu}\psi(x):,
\end{eqnarray}
putting (\ref{final-J})
in (\ref{sugawara in com}) and using the identity (\ref{identity}) for on-shell fermions
 in two dimensions, we find
\begin{eqnarray}
T_{\mu\nu}^{s}=\frac{i}{4}:(\bar\psi\gamma_{\mu}\partial_{\nu}\psi+\bar\psi\gamma_{\nu}\partial_{\mu}\psi
-\partial_{\mu}\bar\psi\gamma_{\nu}\psi-\partial_{\nu}\bar\psi\gamma_{\mu}\psi):,
\end{eqnarray}
 which is
exactly equal to $T_{\mu\nu}$ mentioned in (\ref{com-canonic}).  This equivalence suggests the existence
of a canonical massless pseudo scalar field, satisfying $[\phi(x,t),\dot{\phi}(y,t)]=i\delta(x-y)$, which is
related to the conserved current $j_{\mu}(x)$ through the following equation \cite{Freundlich-2}
 \begin{eqnarray}\label{current-boson}
 j_{\mu}(x)=\frac{1}{\sqrt{\pi}}~\epsilon_{\mu\nu}\partial^{\nu}\phi(x).
 \end{eqnarray}
If we substitute (\ref{current-boson}) into the Sugawara form (\ref{commutative-sugawara}) and use the
identity $\epsilon_{\mu\alpha}\epsilon_{\nu\beta}=g_{\mu\beta}g_{\alpha\nu}-g_{\mu\nu}g_{\alpha\beta}$, it is
found that
 \begin{eqnarray}
T_{\mu\nu}^{s}=\frac{1}{2}\left(\partial_{\mu}\phi\partial_{\nu}\phi+\partial_{\nu}\phi\partial_{\mu}\phi
-g_{\mu\nu}\partial_{\lambda}\phi\partial^{\lambda}\phi \right),
\end{eqnarray}
which describes the energy-momentum tensor for a free massless boson.
Another interesting result of
the Sugawara construction is that it is possible to find explicitly the fermionic filed in terms of the
bosonic field, as mentioned in the introduction part. To show this, we consider the equation describing the transformation of the field $\psi$ under the spatial translation \cite{sommerfield}
 \begin{eqnarray}\label{noether}
 \partial_{x_{1}}\psi(x)=i\bigg[\int
dx'~T_{01}^{s},\psi(x)\bigg],~~~~~P_{0}=\int dx'_{1}~T_{01}^{s},
\end{eqnarray}
where $T_{01}^{s}$ is the Noether current of translational symmetry.
Inserting $T_{01}^{s}$ from (\ref{commutative-sugawara}) into (\ref{noether}), we have
\begin{eqnarray}
\partial_{x_{1}}\psi(x)=\frac{i\pi}{2}\bigg[\int dx'_{1}~\bigg(j_{0}(x')j_{1}(x')+j_{1}(x')j_{0}(x')\bigg),\psi(x)\bigg]_{x_{0}=x'_{0}}.
\end{eqnarray}
  Applying the equal-time commutation relations
\begin{eqnarray}
\bigg[j_{0}(x'),\psi(x)\bigg]_{x_{0}=x'_{0}}=-\psi(x)\delta(x_{1}-x'_{1}),~~~~~~~
\bigg[j_{1}(x'),\psi(x)\bigg]_{x_{0}=x'_{0}}=-\gamma_{5}\psi(x)\delta(x_{1}-x'_{1}),
\end{eqnarray}
 with $\gamma_{5}=\gamma_{0}\gamma_{1}$, we arrive at
 \begin{eqnarray}
\partial_{x_{1}}\psi(x)=-i\pi[j_{1}(x)+j_{0}\gamma_{5}(x)]\psi(x).
\end{eqnarray}
 Solving this equation yields
\begin{eqnarray}\label{solution}
\psi(x)=e^{-i\pi\int_{-\infty}^{x_{1}}  dx'(j_{1}(x')+j_{0}(x')\gamma_{5})}\psi_{0},
\end{eqnarray}
 where $\psi_{0}$ is a constant spinor in space-time. In the final step, we put the
bosonized form of the currents from (\ref{current-boson}) in (\ref{solution})
 \begin{eqnarray}
\psi(x)=e^{i\sqrt{\pi}~[\gamma_{5}\phi(x)+\int_{-\infty}^{x_{1}} dx'_{1} \dot{\phi}(x')]}\psi_{0}.
 \end{eqnarray}
As we see, the spinor field $\psi$ is mapped to the bosonic field $\phi$.
\end{appendix}
 \end{document}